\documentclass[12pt,preprint]{emulateapj}
\usepackage{graphicx}

\shorttitle{Habitable Zone Dependence}
\shortauthors{Stephen R. Kane}
\slugcomment{Submitted for publication in the Astrophysical Journal}

\begin{document}

\title{Habitable Zone Dependence on Stellar Parameter Uncertainties}
\author{Stephen R. Kane}
\affil{Department of Physics \& Astronomy, San Francisco State
  University, 1600 Holloway Avenue, San Francisco, CA 94132, USA}
\email{skane@sfsu.edu}


\begin{abstract}

An important property of exoplanetary systems is the extent of the
Habitable Zone (HZ), defined as that region where water can exist in a
liquid state on the surface of a planet with sufficient atmospheric
pressure. Both ground and space-based observations have revealed a
plethora of confirmed exoplanets and exoplanetary candidates, most
notably from the Kepler mission using the transit detection
technique. Many of these detected planets lie within the predicted HZ
of their host star. However, as is the case with the derived
properties of the planets themselves, the HZ boundaries depend on how
well we understand the host star. Here we quantify the uncertainties
of HZ boundaries on the parameter uncertainties of the host star. We
examine the distribution of stellar parameter uncertainties from
confirmed exoplanet hosts and Kepler candidate hosts and translate
these into HZ boundary uncertainties. We apply this to several known
systems with a HZ planet to determine the uncertainty in their HZ
status.

\end{abstract}



\section{Introduction}
\label{intro}

The discovery of exoplanets has proceeded through a period of
technique refinement over the past 20 years in order to improve
sensitivity to planets of smaller size and larger semi-major axis. As
such, the relevance of the Habitable Zone (HZ) boundaries for
exoplanet host stars has moved from the realm of a theoretical
exercise to one of enormous practical application. A rigorous
calculation of the HZ boundaries for main sequence stars have
previously been provided by \citet{kas93}. These calculations have
since been generalized in terms of stellar luminosity ($L_\odot$) and
effective temperature ($T_\mathrm{eff}$) by a variety of authors
\citep{und03,sel07,jon10}. The relation of the HZ boundaries to
fundamental stellar properties have been recently recalculated and
extended to lower mass stars by \citet{kop13}.

The radial velocity (RV) technique has revealed several possible HZ
terrestrial planets, such as those in the GJ~667 system
\citep{ang13,fer14}. Most of the RV planets in their stars HZ are of
Jovian mass however \citep{kan12}, although this opens up the prospect
of habitable moons in such cases \citep{hin13}. A significant source
of planets which potentially orbit within the HZ of their host stars
has been from the Kepler mission. A recent example of this includes
the Kepler-62 system which contains two planets of $\sim 1.5$ Earth
radii within the HZ \citep{bor13}. A point of caution though lies
within the uncertainty on the stellar parameters on which the HZ
boundaries sensitively depend. In some cases, this can lead to a
radical reevaluation of the HZ which can change the HZ status of
planets within the system \citep{kal11,man13}. Thus it is important to
study in detail how the current HZ estimates are impacted by the
uncertainty in the host star properties.

Here we describe the fundamental stellar properties which are used to
determine the HZ boundaries and quantify the effect of stellar
parameters uncertainties on HZ calculations. Section 2 includes a
summary of the methodology used to determine the HZ boundaries
throughout the rest of this work. In Section 3, we examine each of the
primary contributing stellar parameters in detail with respect to how
the uncertainties translate to our understanding the extent of the
HZ. These results are applied to two populations of exoplanet host
stars in Section 4, broadly defined as the ``confirmed exoplanet host
stars'' and the ``Kepler candidate host stars''. In Section 5, we
examine several specific cases of claimed HZ exoplanets to determine
the validity of those claims and provide concluding remarks in Section
6.


\section{The Habitable Zone Boundaries}
\label{hz}

\begin{figure*}
  \begin{center}
    \begin{tabular}{cc}
      \includegraphics[angle=270,width=8.2cm]{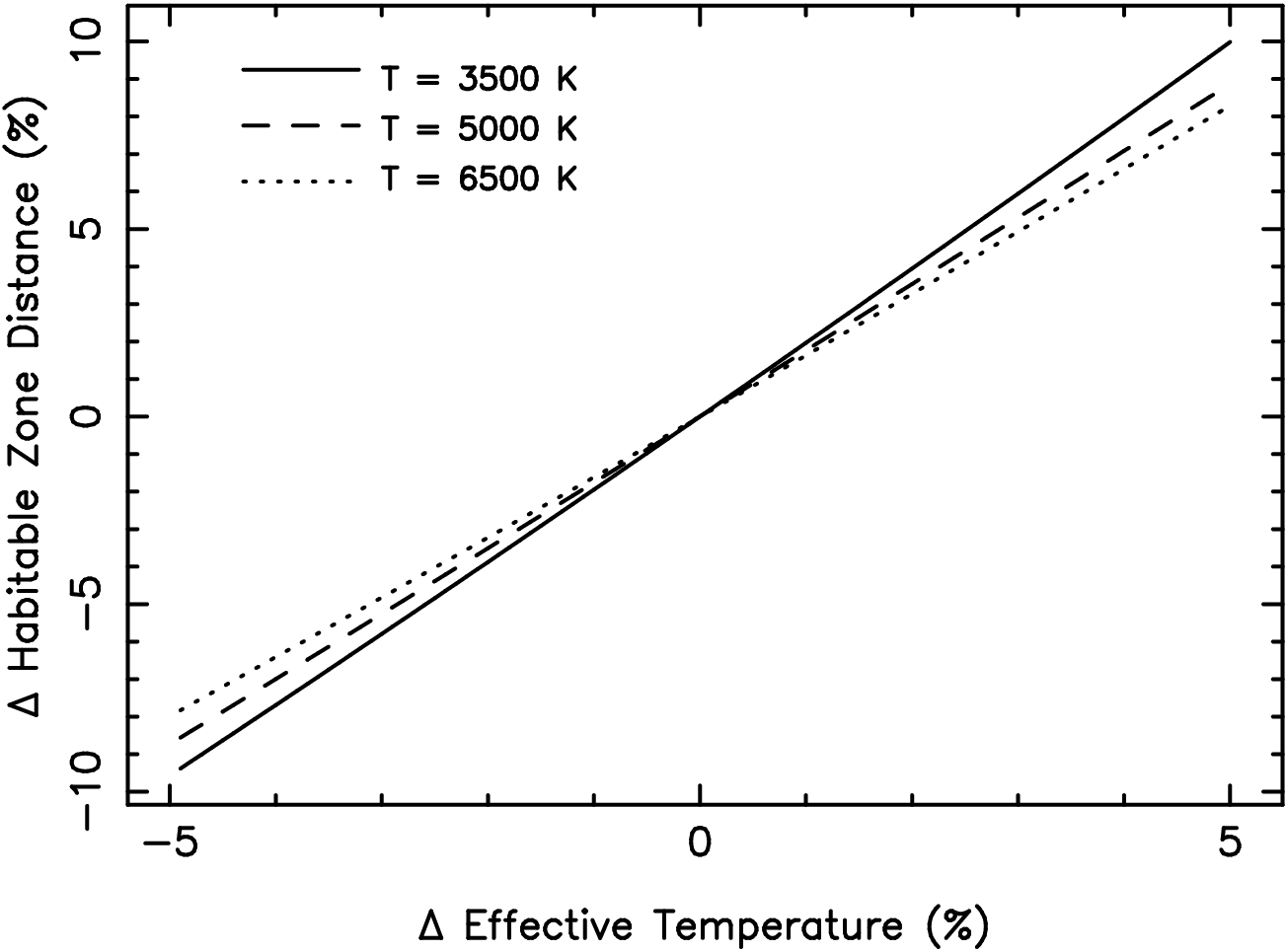} &
      \includegraphics[angle=270,width=8.2cm]{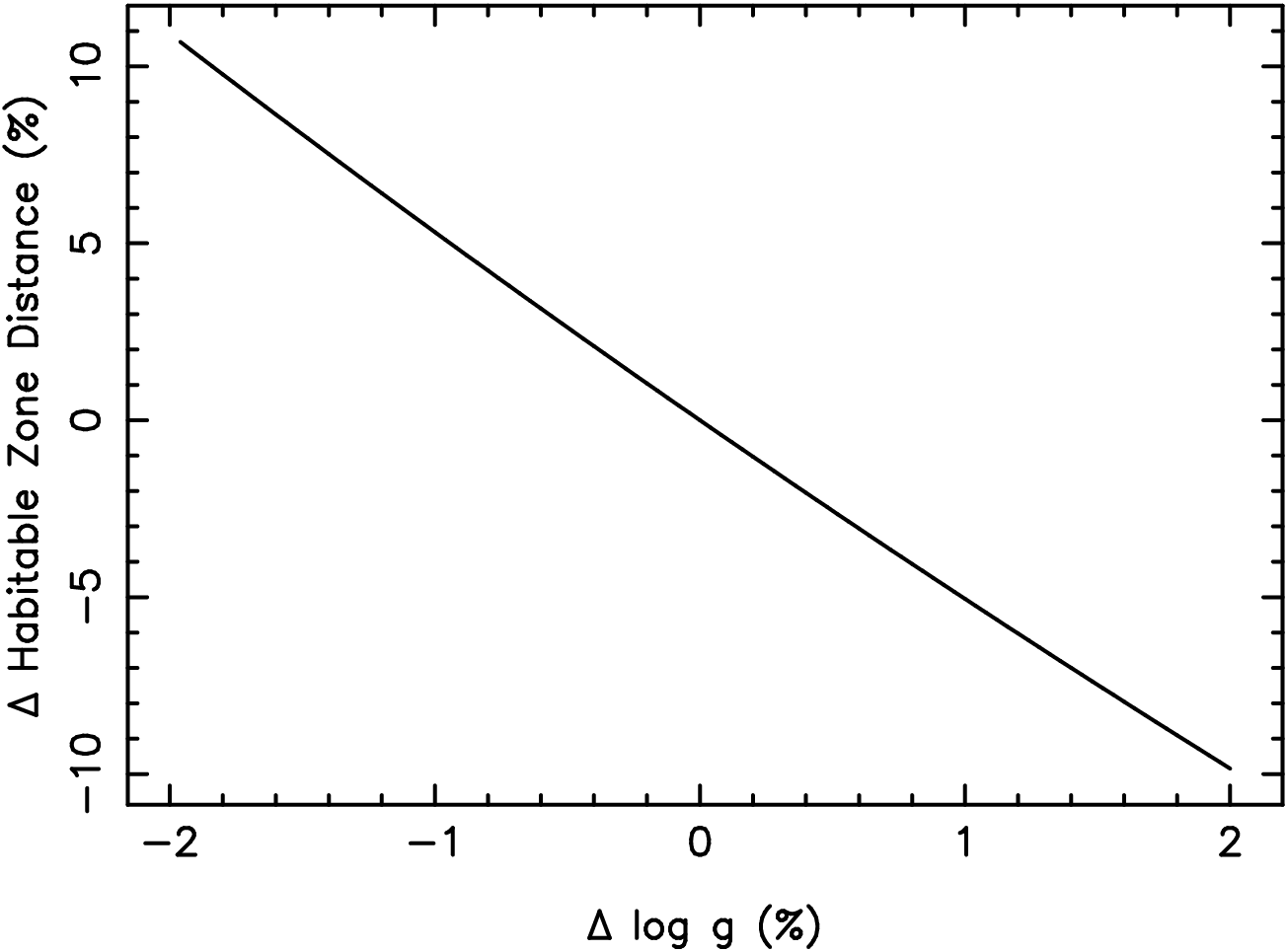}
    \end{tabular}
  \end{center}
  \caption{The dependency of the ``Runaway Greenhouse'' HZ boundary
    (see Section \ref{hz}) on the percentage changes in stellar
    effective temperature (left) and surface gravity (right). Note
    that the dependency on surface gravity arises when used to
    determine the stellar radius and thus stellar luminosity.}
  \label{paramuncfig}
\end{figure*}

The HZ for a particular star is broadly defined as the region around
the star where water can exist in the liquid state on the surface of a
planet given adequate atmospheric pressure. These boundaries were
calculated by \citet{kas93} based upon one-dimensional, cloud-free
climate models. The most recent revision of these calculations by
\citet{kop13} uses updated models for H$_2$O and CO$_2$ in the
thermal-IR and expands the range of stellar effective temperatures for
which the calculations are valid. A generalization of the incident
stellar flux at the HZ boundaries is as follows:
\begin{equation}
  S_\mathrm{eff} = S_\mathrm{eff\odot} + a T_\star + b T_\star^2 + c
  T_\star^3 + d T_\star^2
  \label{stellarflux}
\end{equation}
where $T_\star = T_\mathrm{eff} - 5780$~K and $T_\mathrm{eff}$ is the
stellar effective temperature. \citet{kop13} provide values for the
coefficients $S_\mathrm{eff\odot}$, $a$, $b$, $c$, and $d$ depending
on assumptions on when water-loss may have occurred in the Venusian
and Martian history. Here we will refer to ``conservative'' and
``optimistic'' models of the HZ, as used by \citet{kan13}. The
conservative model treats the ``Runaway Greenhouse'' and ``Maximum
Greenhouse'' criteria as the inner and outer HZ boundaries
respectively. The optimistic model adopts the ``Recent Venus'' and
``Early Mars'' criteria for these inner and outer boundaries, allowing
for an expanded HZ under the assumption that Venus and Mars may have
had a longer period of retaining surface water. These criteria are
described in more detail by \citet{kop13}. The distance, $d$, of the
boundaries for a particular HZ model may be determined from the
stellar flux in Equation \ref{stellarflux} using the following
relation:
\begin{equation}
  d = \left( \frac{L_\star/L_\odot}{S_\mathrm{eff}}
  \right)^{0.5}
  \label{hzbound}
\end{equation}
where $L_\star$ is the stellar luminosity and the distance is in units
of AU.

The above calculations depend sensitively on $T_\mathrm{eff}$ and
$L_\star$. The luminosity of the nearest stars may be determined
through the use of stellar parallax and subsequent distance
estimates. We discuss the uncertainties associated with such
luminosity determinations in Section \ref{paramunc}. Distance
estimates of reasonable accuracy are in limited supply for many
exoplanet host stars and the luminosity is often calculated based on
assumptions regarding the stellar radius, $R_\star$, via the equation
$L_\star = 4 \pi R_\star^2 \sigma T_\mathrm{eff}^4$. The stellar
radius is often calculated based on stellar models and the relatively
accessible quantities of $T_\mathrm{eff}$ and surface gravity, $\log
g$, since these influence the stellar photosphere. The radius may also
be estimated from the stellar mass, $M_\star$, and surface
gravity. Below is an example relationship \citep{sma05} which allows
the radius to be determined from other fundamental stellar properties:
\begin{equation}
  \log g = \log \left( \frac{M_\star}{M_\odot} \right) - 2 \log
  \left( \frac{R_\star}{R_\odot} \right) + \log g_\odot
  \label{logg}
\end{equation}
where $\log g_\odot = 4.4374$. The determination of the HZ boundaries
can thus be fraught with uncertainty depending on how well the stellar
properties are known.


\section{Stellar Parameter Uncertainties}
\label{paramunc}

The two main quantities used to determine the various HZ boundaries
are $T_\mathrm{eff}$ and $L_\star$. As seen in Equation \ref{hzbound},
the boundaries have a power law dependence on $L_\star$. However,
since $L_\star$ is a calculated rather than measured quantity, it
depends upon other variables ($T_\mathrm{eff}$, $R_\star$, and $\log
g$), each with their own uncertainties. Here we quantify the HZ
boundary dependencies on these parameters.

The first parameter we consider is that of effective temperature. As
seen in Equation \ref{stellarflux}, this parameter is fundamental to
calculating the stellar flux received at the HZ
boundaries. Fortunately, it is also usually one of the better known
stellar parameters with direct measurements from stellar spectra. As
such, it has little dependency upon the knowledge of other stellar
parameters. The uncertainty in $T_\mathrm{eff}$ also filters through
to the determination of luminosity since that is often calculated from
$T_\mathrm{eff}$ and $R_\star$. Thus, the $T_\mathrm{eff}$ uncertainty
plays a major role in determining the robustness of the HZ boundaries.

The left panel of Figure \ref{paramuncfig} shows the dependency of the
``Runaway Greenhouse'' boundary (inner conservative HZ boundary) on
percentage uncertainties in $T_\mathrm{eff}$. This shows that
uncertainties in $T_\mathrm{eff}$ of $\sim 5$\% can result in $\sim
10$\% uncertainties in the location of the HZ boundary. There is a
variation of this dependency on spectral type but the variation is
relatively minor in nature.

The dependency of HZ boundary determinations on $T_\mathrm{eff}$ is
independent of the stellar radius, which is usually a calculated
rather than measured quantity. However, the uncertainty in $R_\star$
contributes significantly to the HZ error budget and is often
relatively large. There is a linear dependence of the HZ distances on
$R_\star$ since $L_\star \propto R_\star^2$ and $d \propto
\sqrt{L_\star}$ (see Equation \ref{hzbound}). Thus the uncertainty in
the location of the HZ depends linearly on the $R_\star$
uncertainties.

\begin{figure*}
  \begin{center}
    \begin{tabular}{cc}
      \includegraphics[angle=270,width=8.2cm]{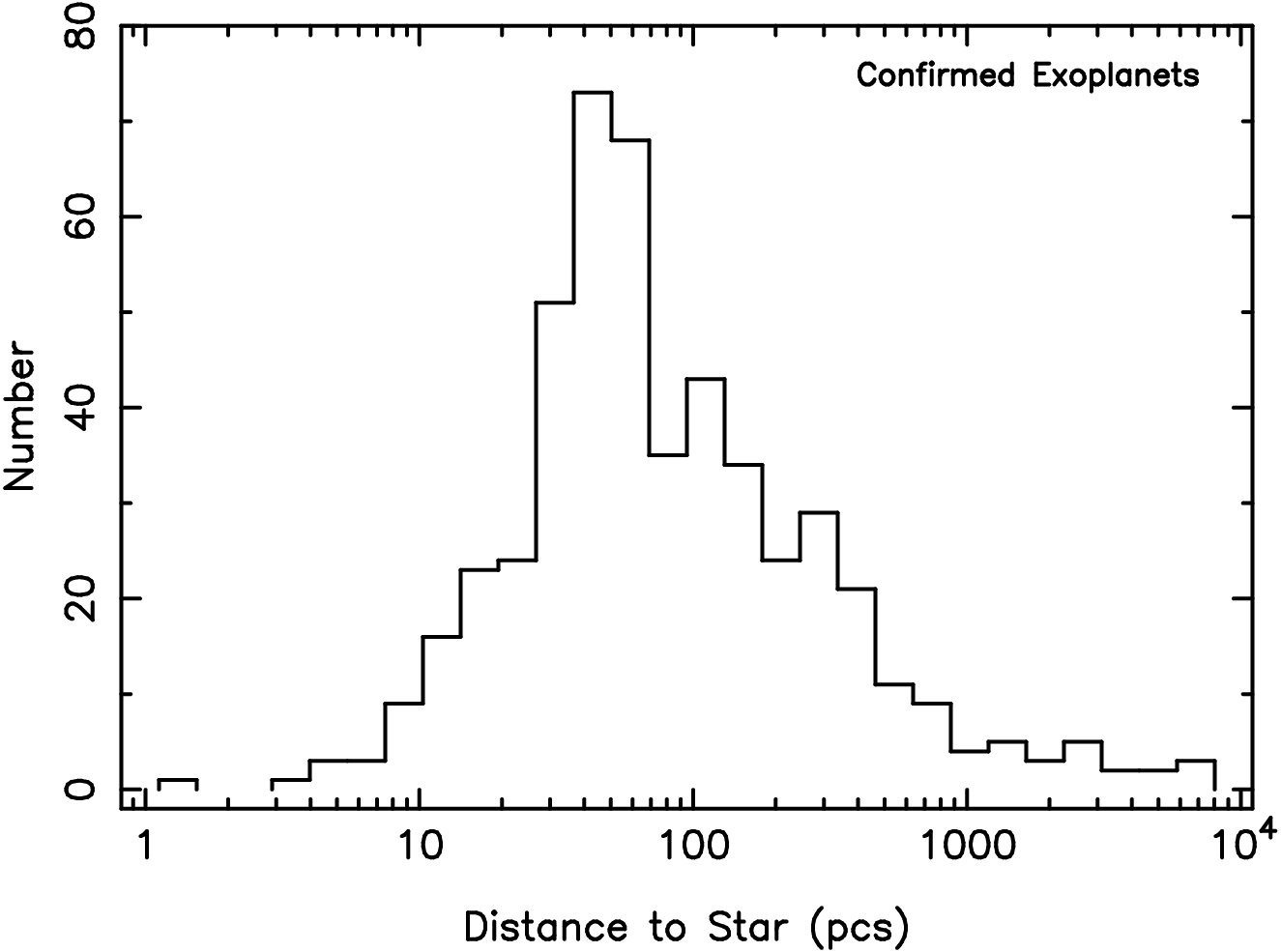} &
      \includegraphics[angle=270,width=8.2cm]{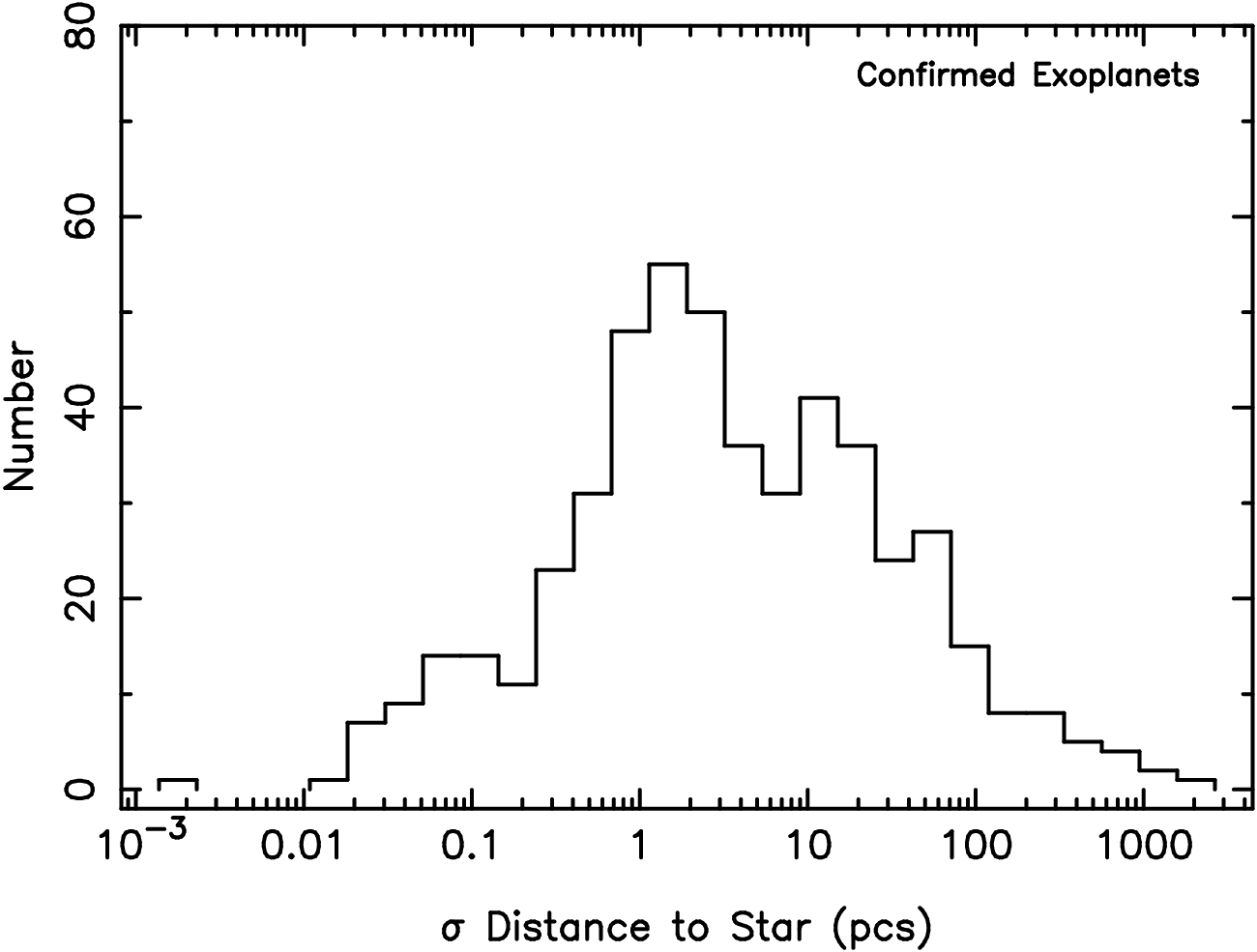}
    \end{tabular}
  \end{center}
  \caption{Histograms of the stellar distances (left panel) and their
    associated uncertainties (right panel) for the confirmed
    exoplanet host stars as determined from their stellar
    parallax. Each histogram uses a total of 30 bins.}
  \label{distfig}
\end{figure*}

We also consider the contribution of the surface gravity ($\log g$)
and stellar mass ($M_\star$) uncertainties to the error budget since
these can be used to determine $R_\star$. According to Equation
\ref{logg}, the dependency of $R_\star$ on $\log g$ and $M_\star$ are
$R_\star \propto \sqrt{1/g}$ and $R_\star \propto \sqrt{M_\star}$
respectively. Thus the HZ boundaries have the same dependencies: $d
\propto \sqrt{1/g}$ and $d \propto \sqrt{M_\star}$. The right panel of
Figure \ref{paramuncfig} plots the effect of varying $\log g$ on the
HZ boundaries which shows the inverse dependence. Note that the
dependencies may be more complicated than described since a change in
mass for a main sequence star will result in a change in radius, thus
an associated change in $g$.

Finally, we consider luminosity determinations from stellar distances
derived from their parallax measurements. The parallax measurements
are typically those provided by the Hipparcos mission \citep{per97}
and the revised reduction of those data \citep{van07}. This is a
common method for luminosity determination for the confirmed
exoplanets whose discovery resulted from the RV technique. In these
cases the stars tend to be preferentially bright and thus close enough
for reasonable parallax measurements. Figure \ref{distfig} shows
histograms of distance measurements and their associated uncertainties
for the confirmed exoplanet host stars. These data were extracted from
the Exoplanet Data Explorer (see Section \ref{application}). The vast
majority of RV host stars are closer than 100 pcs where distances are
determined to $\pm$~1--10\%. The luminosity is proportional to the
square of the distance and also depends on the quality of the
photometry and bolometric corrections. As such, luminosity
determinations using this technique also have the potential for
relatively large uncertainties for all but the closest stars. This
situation will undoubtedly be greatly improved by new astrometric
measurements provided by the GAIA mission\footnote{\tt
  http://sci.esa.int/gaia/}. In the following sections, we restrict
ourselves to using the stellar parameters noted above in order to
provide a direct comparison between different types of exoplanet host
stars. The caveat to note is that some of the confirmed exoplanet host
stars may have overestimated HZ boundary uncertainties where their
distances are suffiently well known.


\section{Application to Known Exoplanet Host Stars}
\label{application}

\begin{figure*}
  \begin{center}
    \begin{tabular}{cc}
      \includegraphics[angle=270,width=8.2cm]{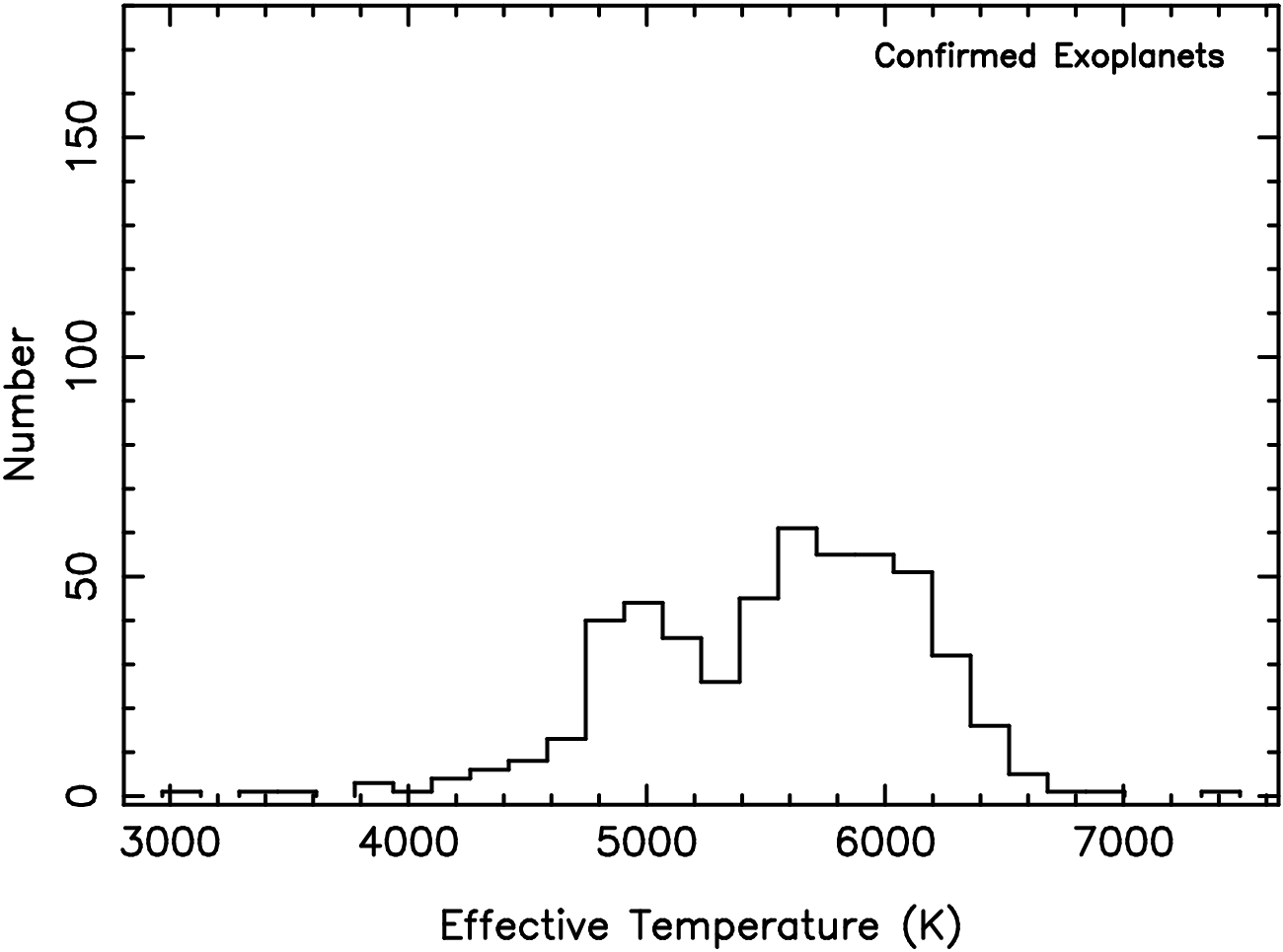} &
      \includegraphics[angle=270,width=8.2cm]{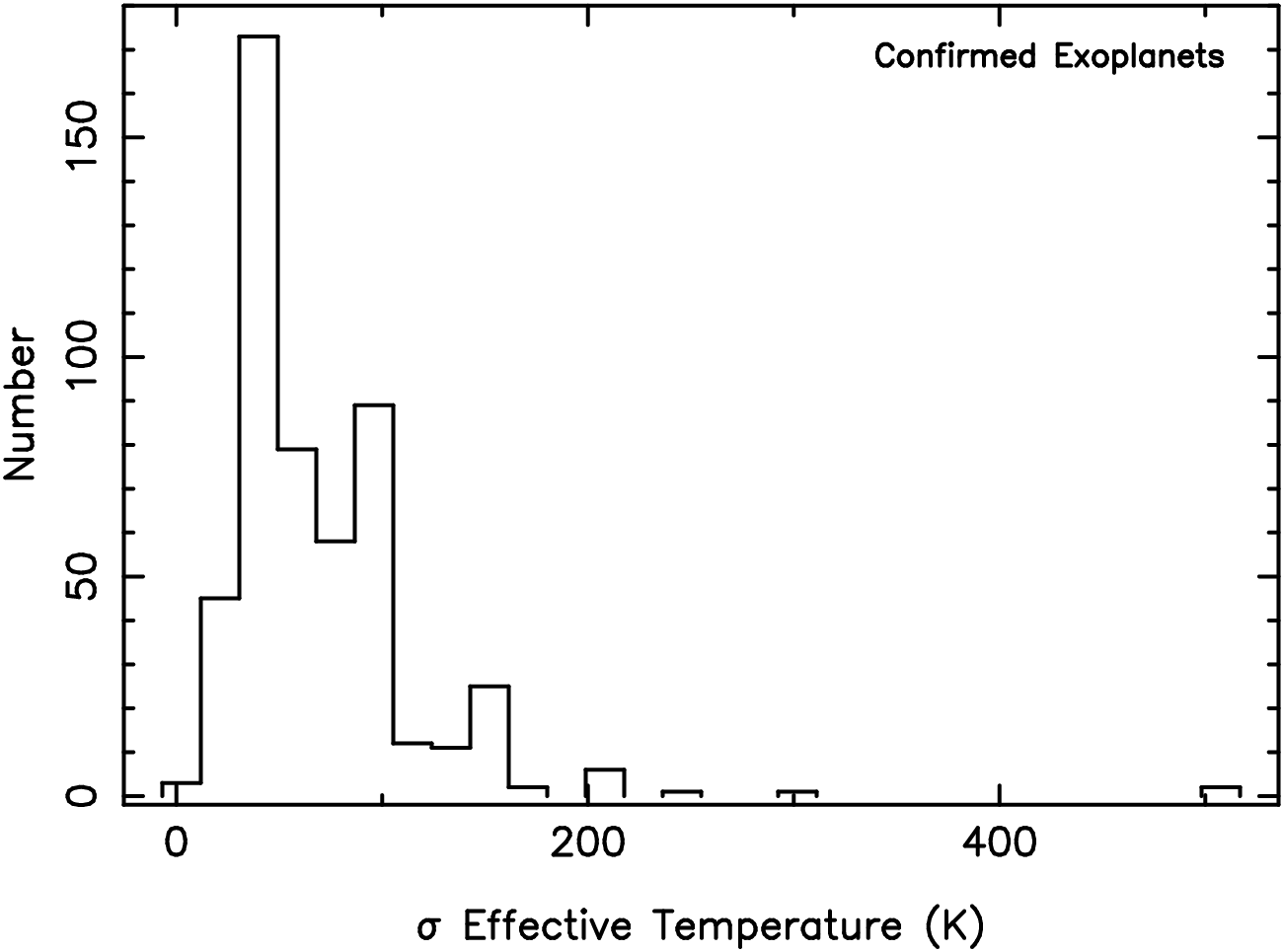} \\ \\ \\
      \includegraphics[angle=270,width=8.2cm]{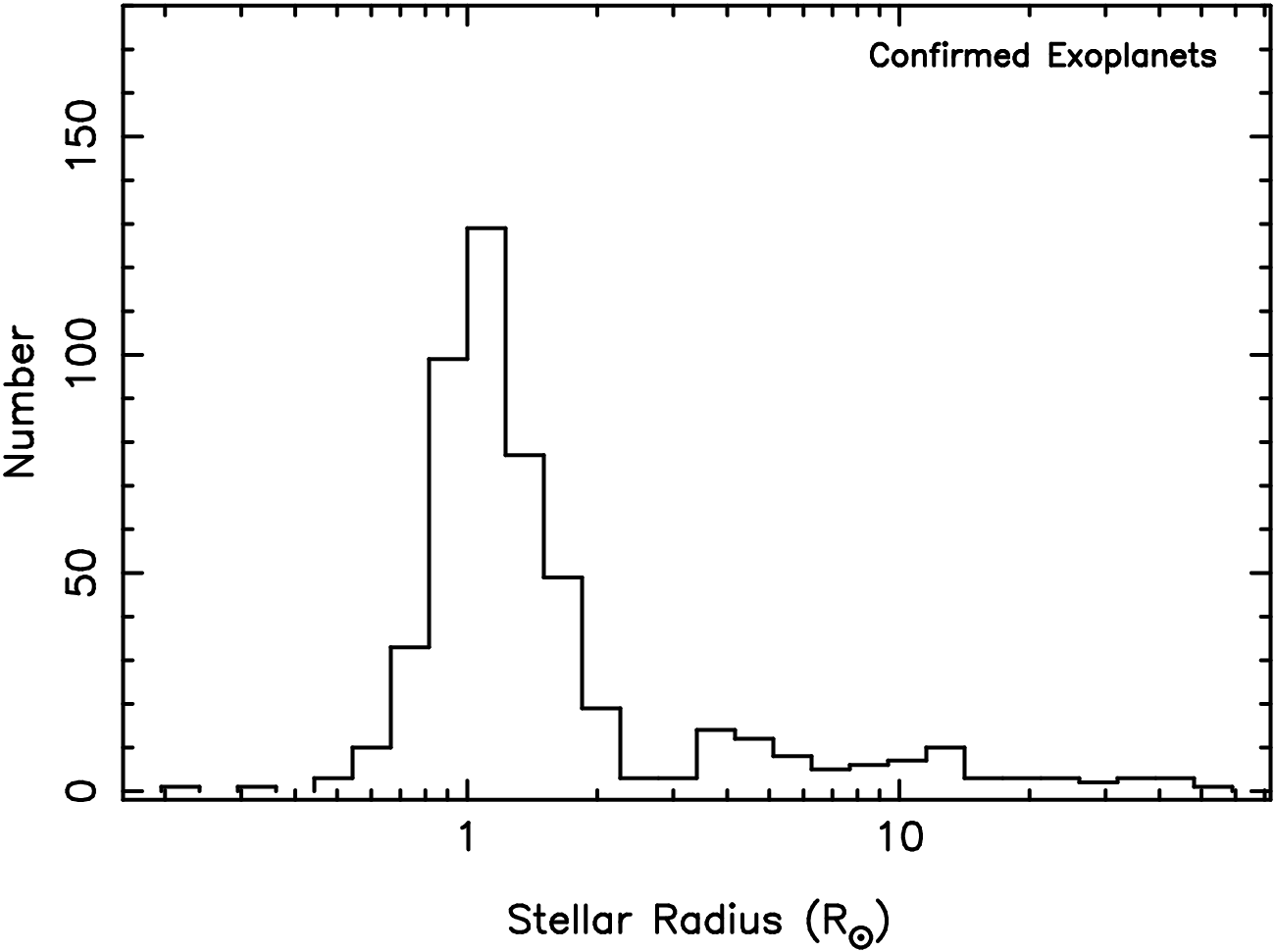} &
      \includegraphics[angle=270,width=8.2cm]{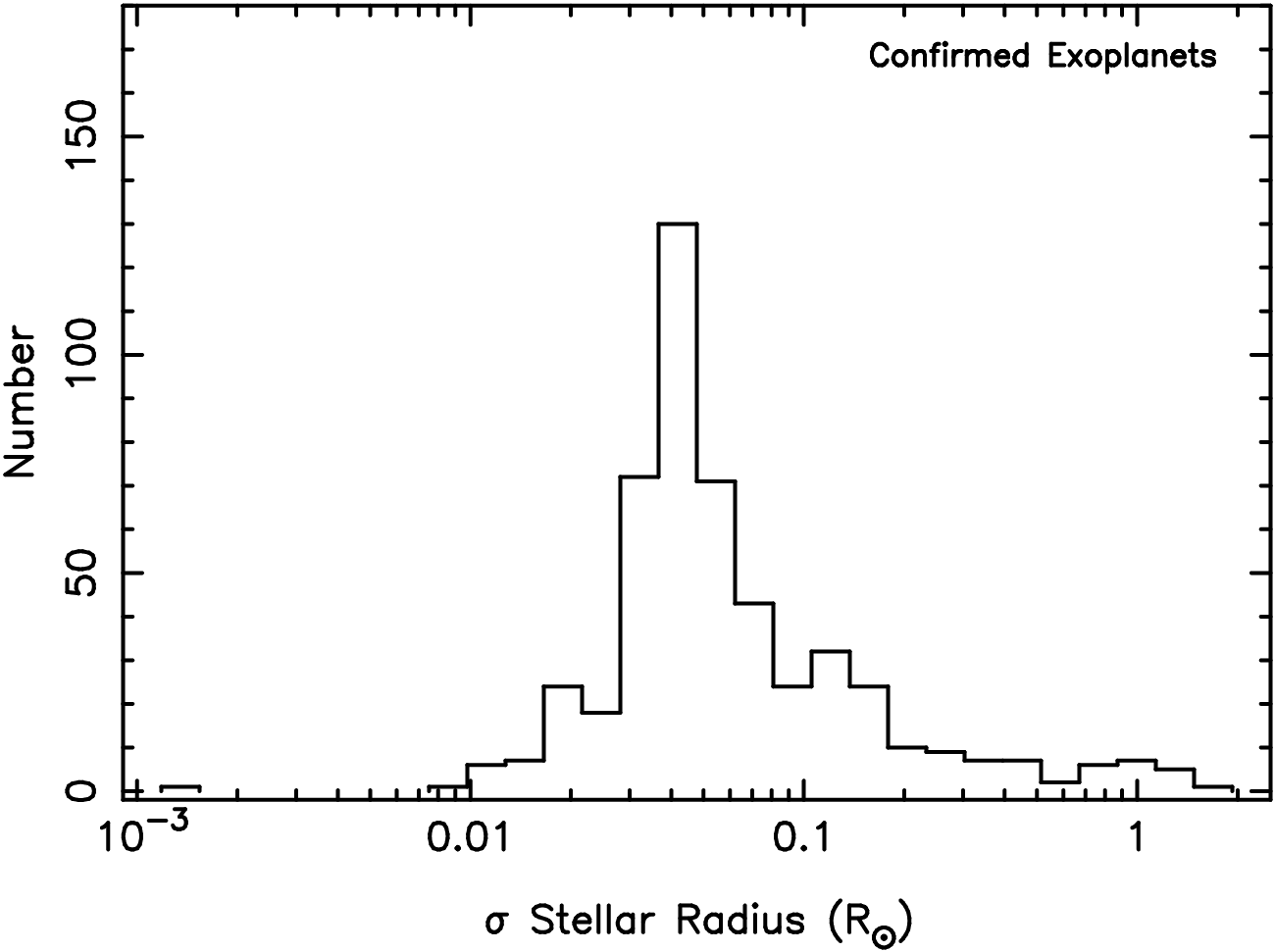} \\ \\ \\
      \includegraphics[angle=270,width=8.2cm]{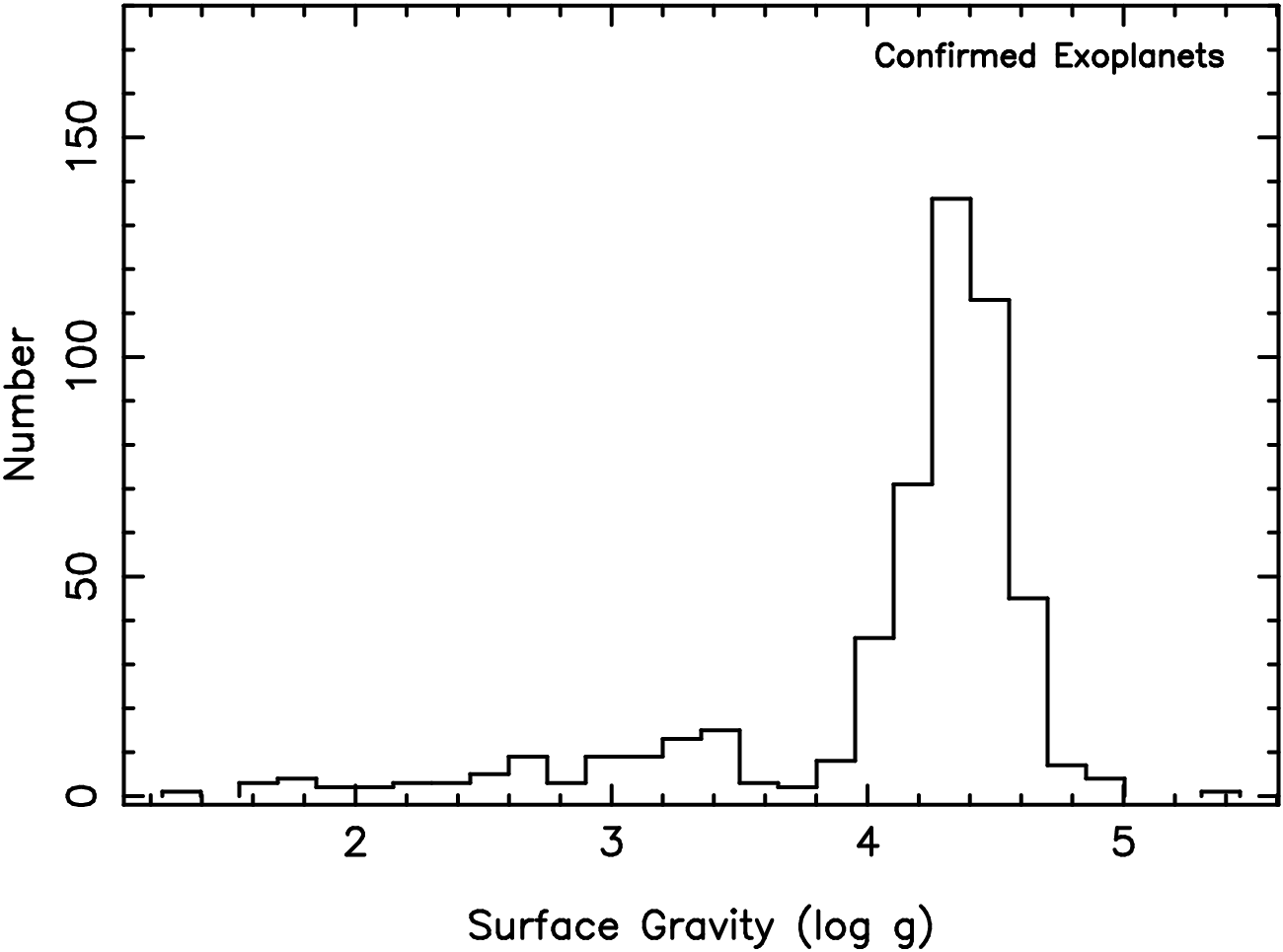} &
      \includegraphics[angle=270,width=8.2cm]{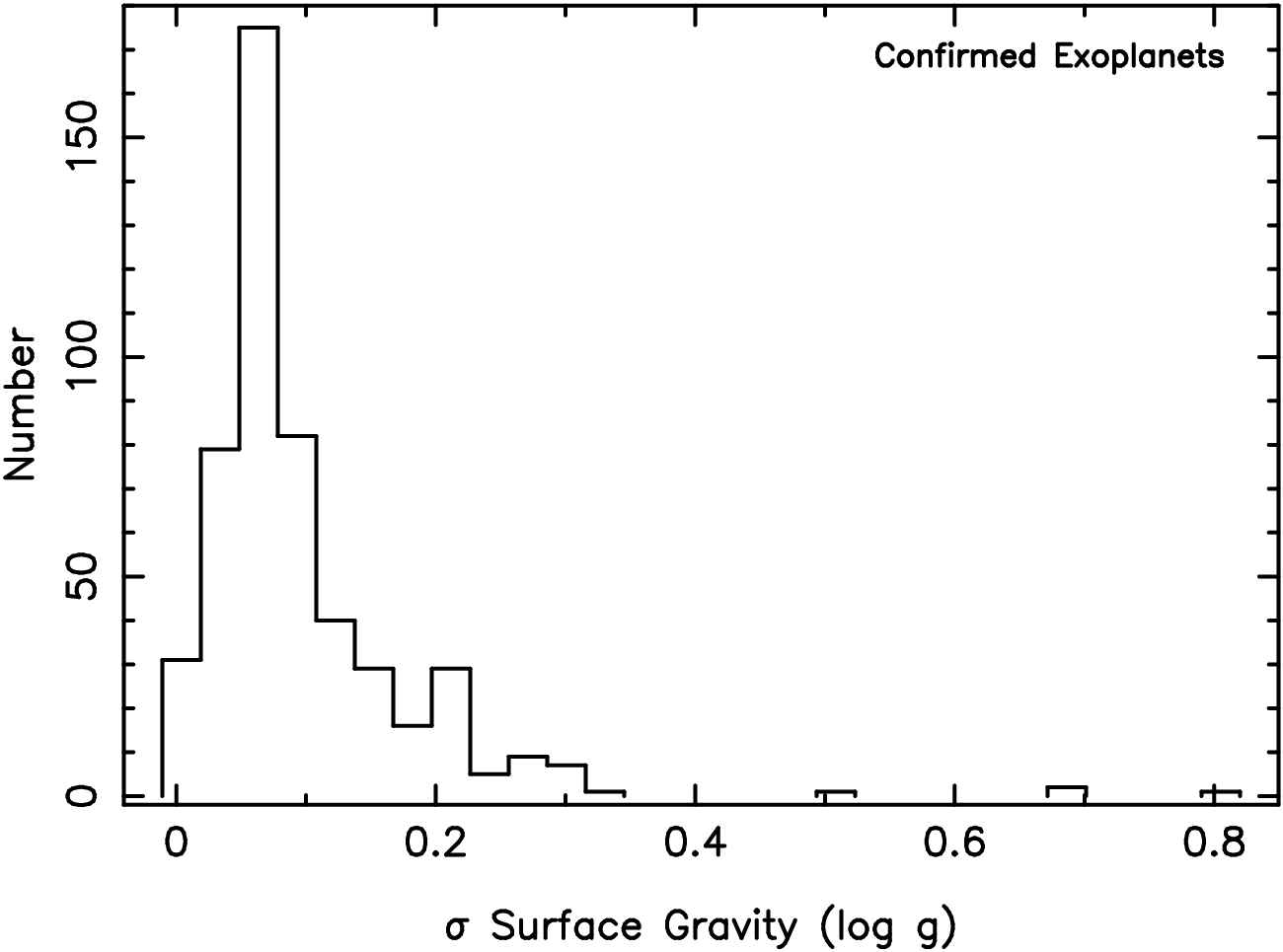}
    \end{tabular}
  \end{center}
  \caption{Histograms of the stellar parameters (left column) and
    their associated uncertainties (right column) for the confirmed
    exoplanets group. Included parameters are effective temperature
    (top row), stellar radius (middle row), and surface gravity
    (bottom row). Each histogram uses a total of 30 bins with data for
    507 host stars. Note that the stellar radius histograms have a
    logarithmic scale.}
  \label{cpparams}
\end{figure*}

\begin{figure}
  \includegraphics[angle=270,width=8.2cm]{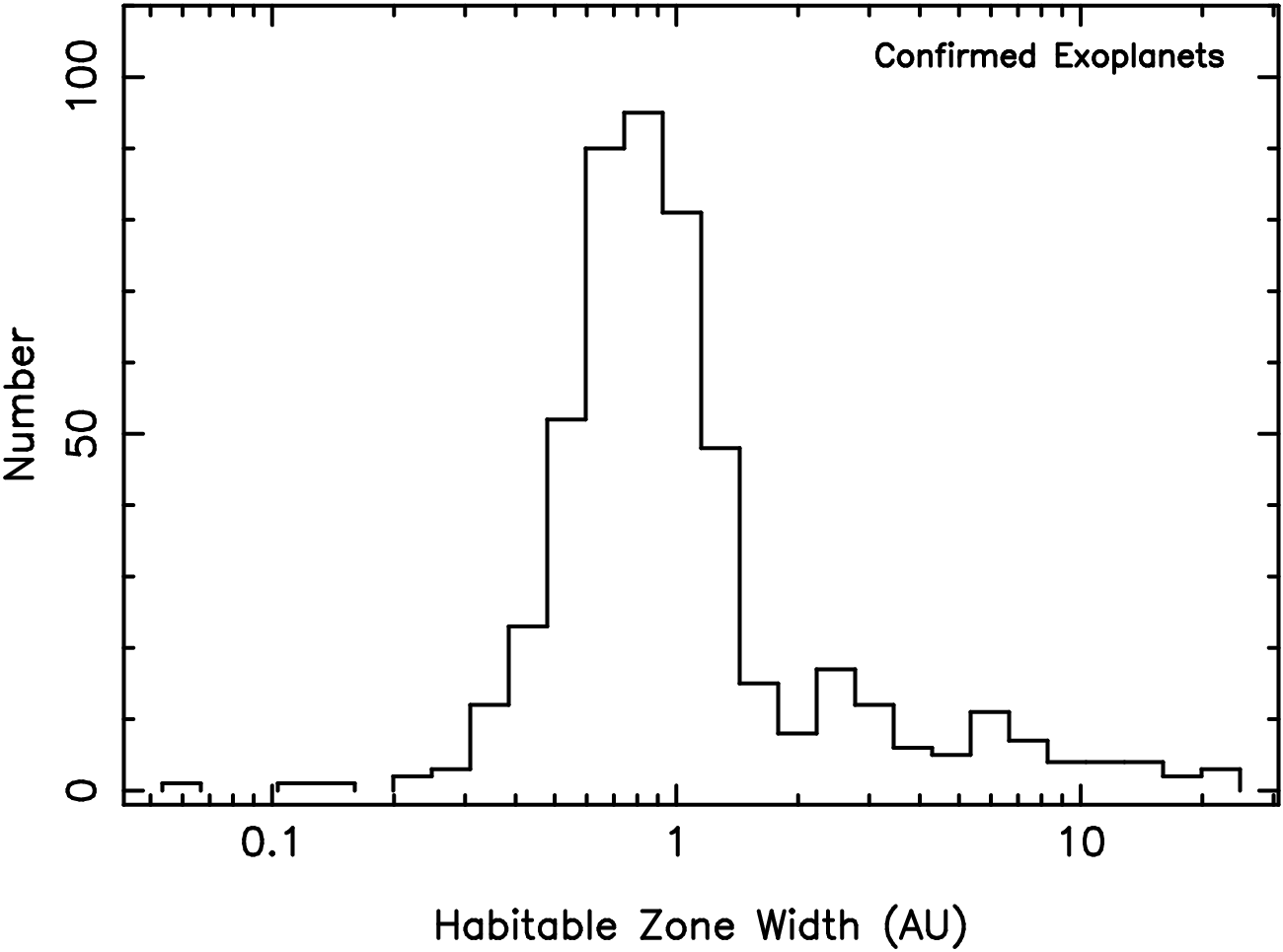} \\
  \includegraphics[angle=270,width=8.2cm]{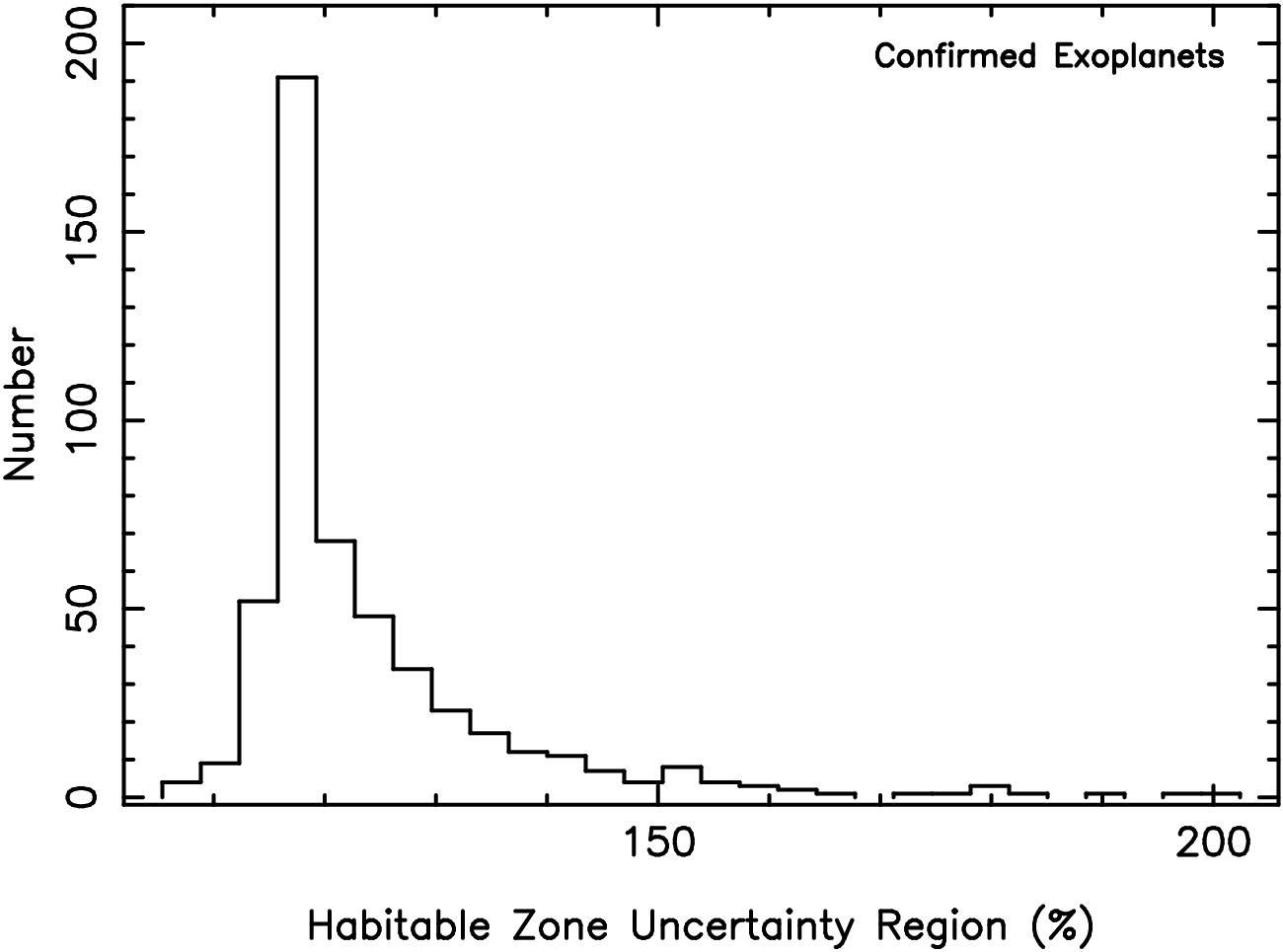}
  \caption{Histogram representations of the HZ distributions for the
    confirmed exoplanets group. Top: Histogram of the calculated HZ
    width (AU). Bottom: Histogram of the width of the HZ uncertainty
    region, calculated as a percentage of the HZ width shown in the
    top panel.}
  \label{cphz}
\end{figure}

Here we apply the above described stellar parameter uncertainty
effects to the HZ calculations for the known exoplanet host stars. We
divide these into two broad groups: the host stars of confirmed
exoplanets and the host stars of Kepler candidates. We extracted the
stellar parameters stored in the Exoplanet Data Explorer\footnote{\tt
  http://exoplanets.org/} \citep{wri11}. The data are current as of
26th November 2013. The data are also available from the NASA
Exoplanet Database\footnote{\tt
  http://exoplanetarchive.ipac.caltech.edu/} \citep{ake13}. The data
utilized here includes all of the host stars with the necessary
stellar parameter information to perform this analysis. This main
criteria are that the host stars have available values for the
effective temperatures, radii, and surface gravities.

An inherent assumption in the following analysis is that the
uncertainties associated with stellar parameters are gaussian in
nature and can mapped to 1$\sigma$ uncertainties in HZ boundary
locations. There are numerous individual cases where this will not be
true due to non-gaussian posteriors and correlated errors which are
inherent in the analysis of spectra (for example). This does not have
a significant effect on the result since we draw upon a large
distribution to show the impact of these stellar uncertainties.


\subsection{Confirmed Exoplanet Host Stars}

The number of host stars in the confirmed exoplanets group of host
stars which meet the criteria of available stellar parameters is 507.
This group includes host stars for which their planets have been
detected by either the RV or transit techniques. Many of the stellar
properties have been compiled by such sources as \citet{but06} and
\citet{tak07}. Figure \ref{cpparams} summarizes the stellar properties
of $T_\mathrm{eff}$, $R_\star$, and $\log g$. The histograms show the
distribution of the parameters and their associated
uncertainties. Each of the histograms are divided into 30 bins and
have the same y-axis scale for ease of comparison.

The distribution of $T_\mathrm{eff}$ lies mostly between 4000--6500~K;
a reflection of the F--G--K target selection which dominates RV
exoplanet host stars. The distribution of the $T_\mathrm{eff}$
uncertainties peaks at $\sim 50$~K which is $\sim 1$\% of the typical
$T_\mathrm{eff}$ value.

As described in Section \ref{paramunc}, the stellar radius plays a key
role in determining the uncertainty in the HZ boundaries. The stellar
radii histogram shows that dwarf stars dominate as hosts of the
confirmed exoplanets with a distribution peaking at $\sim 1
R_\odot$. The uncertainty distribution peaks at $\sim 0.04 R_\odot$
and thus the typical radius uncertainty for this group of host stars
is $\sim 4$\%.

The distribution of $\log g$ values shown in Figure \ref{cpparams} is
also indicative of the dominance of dwarf host stars in the sample
since it peaks between 4--5. The uncertainty in $\log g$ peaks at
$\sim 0.07$. The range of uncertainty values is sufficient to
determine a luminosity class, but can result in substantial ambiguity
in derived stellar radius when Equation \ref{logg} is employed.

From the values of $T_\mathrm{eff}$ and $R_\star$ described above, we
calculated the HZ boundaries for the conservative model (see Section
\ref{hz}). The width of the conservative HZ was then determined for
each of the stars. The distribution of the HZ widths is represented by
the histogram shown in the top panel of Figure \ref{cphz}. This
distribution appears to be symmetric but is shown on a logarithmic
scale and peaks at a width of $\sim 0.9$~AU. To examine the effects of
the stellar parameter uncertainties, we calculated the HZ boundaries
of the conservative HZ model (see Equations \ref{stellarflux} and
\ref{hzbound}) by subtracting and adding the 1$\sigma$ parameter
uncertainties for the inner and outer boundaries respectively. This
results in an uncertainty region which is larger than the calculated
width of the HZ. The distribution of this HZ uncertainty region is
shown in the bottom panel of Figure \ref{cphz} as a percentage of the
HZ width. Thus, an uncertainty region close to 100\% means that the
stellar parameter uncertainties are relatively small and the location
of the HZ is well constrained. An uncertainty region of 200\% means
that any planet in the HZ of such a system may not be in the HZ at the
$1\sigma$ level. The distribution shown in Figure \ref{cphz} shows
that typical HZ uncertainty regions are only 20\% larger than the
calculated HZ width for the confirmed exoplanet group.


\subsection{Kepler Candidate Host Stars}

There have been several releases of Kepler candidates which is an
ever-growing list of likely transiting exoplanets detected by the
Kepler mission \citep{bor11a,bor11b,bat13}. With the number of Kepler
candidate host stars numbering in their thousands, determining
accurate stellar parameters is a daunting task. Stellar parameters for
these stars have been measured and estimated using a combination of
photometric calibration, spectroscopy, and astroseismology
\citep{bro11,sil12,eve13,hub13}. There are numerous difficulties in
determining these stellar properties which can result in systematic
offsets in parameters such as temperature, matallicity, radius,
etc. These are described in detail by \citet{hub14} and references
therein. This highlights the need for accurate stellar parameters in
order to decrease the HZ boundary uncertainties discussed here.

\begin{figure*}
  \begin{center}
    \begin{tabular}{cc}
      \includegraphics[angle=270,width=8.2cm]{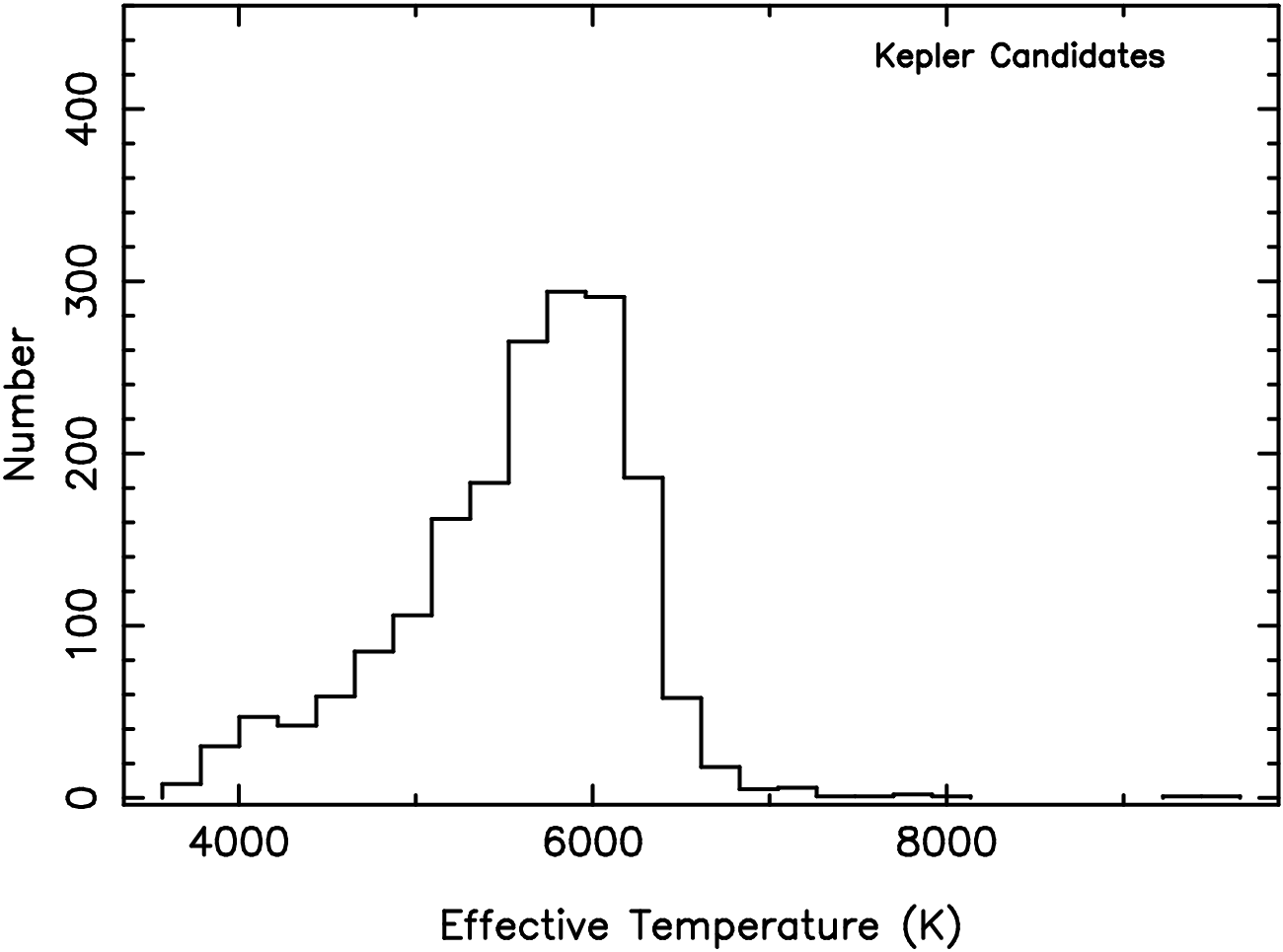} &
      \includegraphics[angle=270,width=8.2cm]{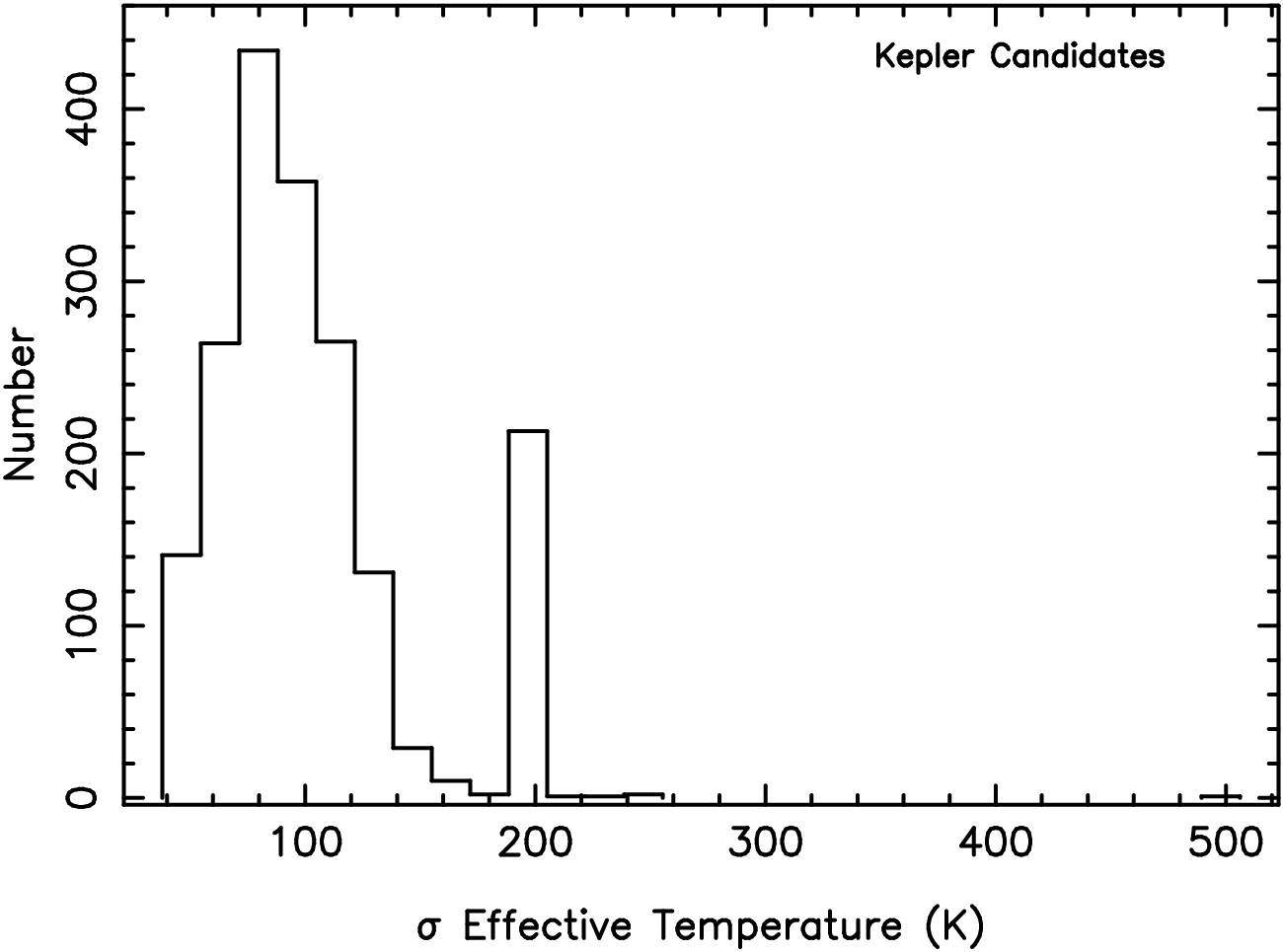} \\ \\ \\
      \includegraphics[angle=270,width=8.2cm]{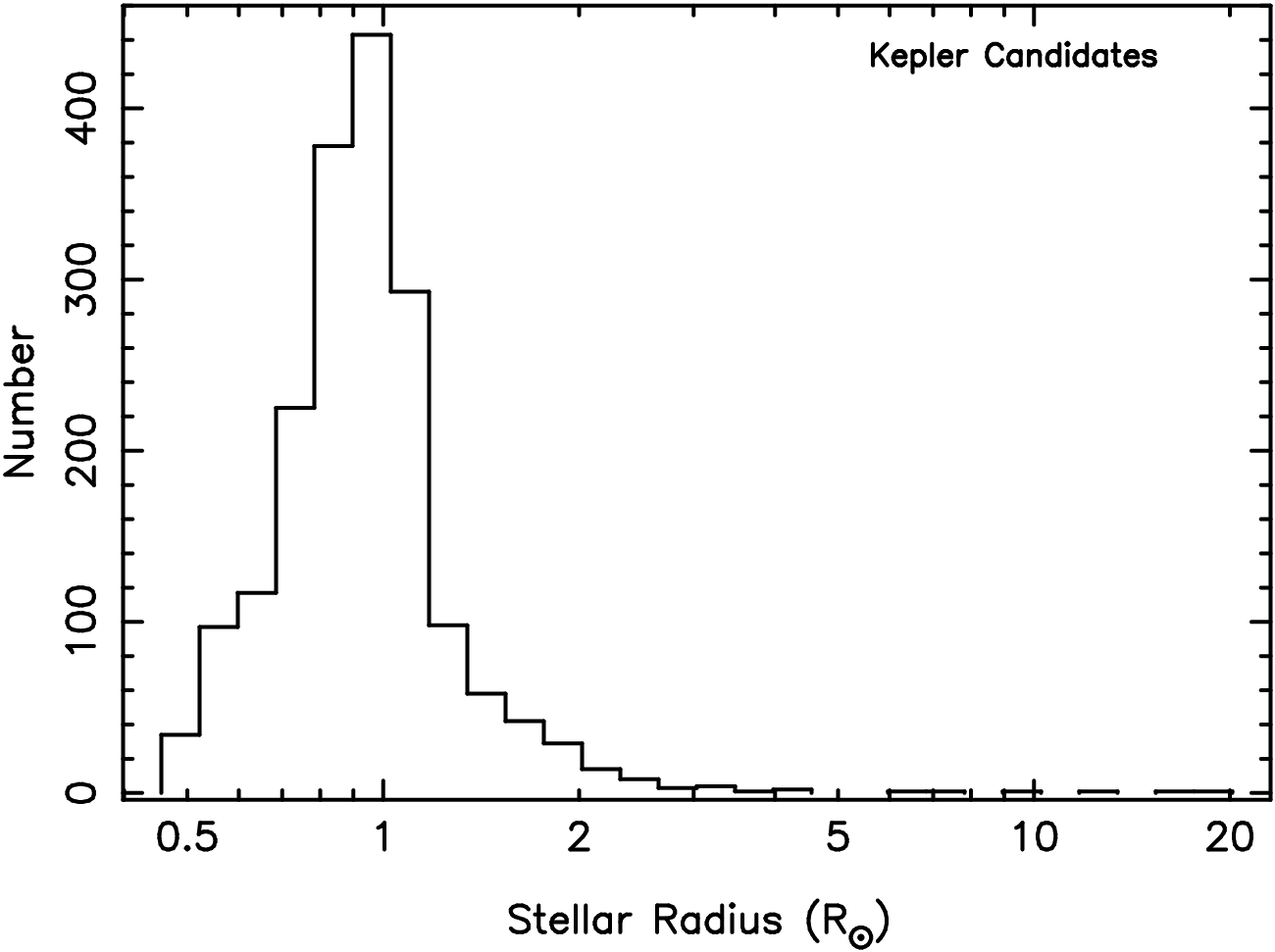} &
      \includegraphics[angle=270,width=8.2cm]{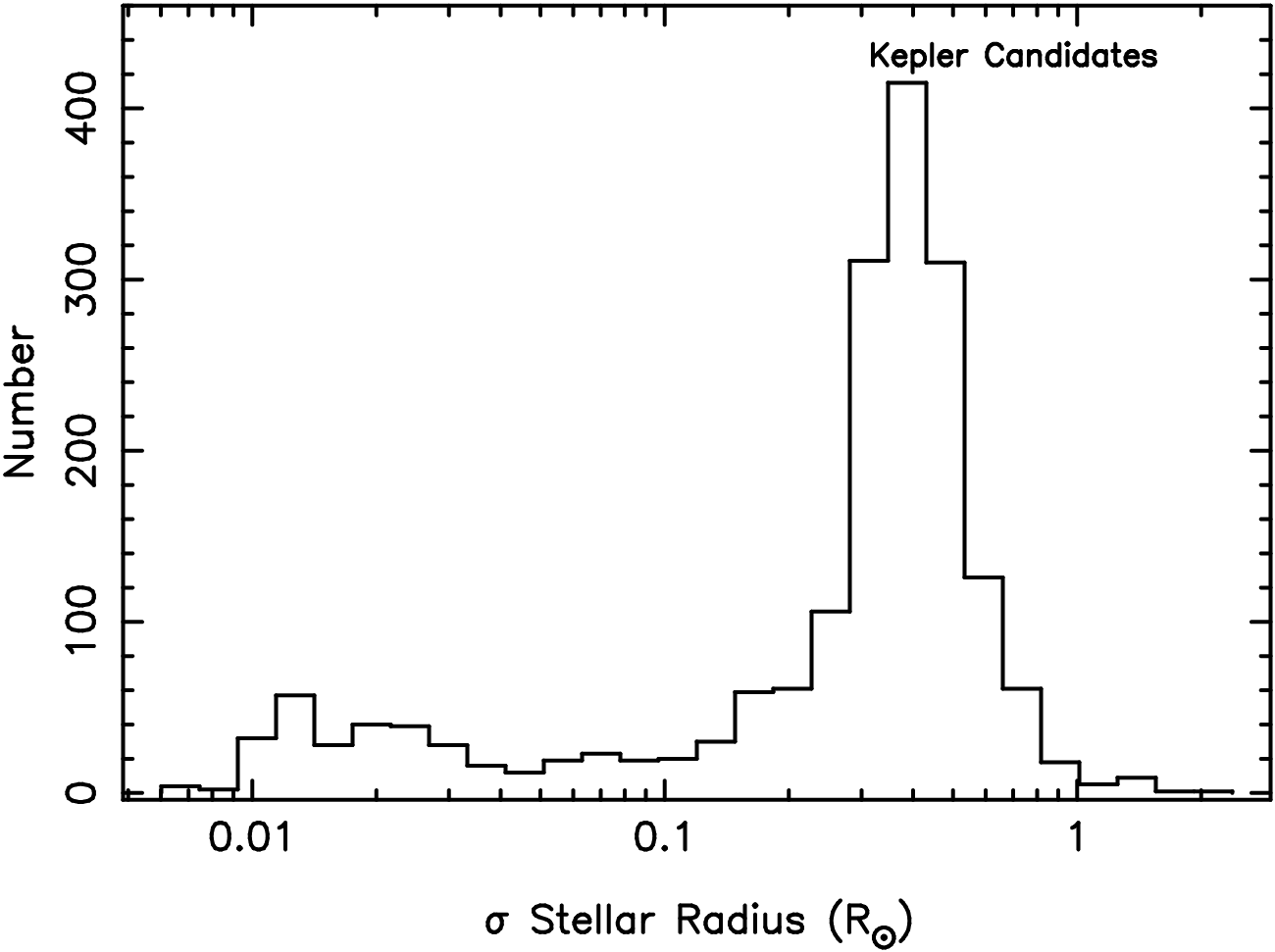} \\ \\ \\
      \includegraphics[angle=270,width=8.2cm]{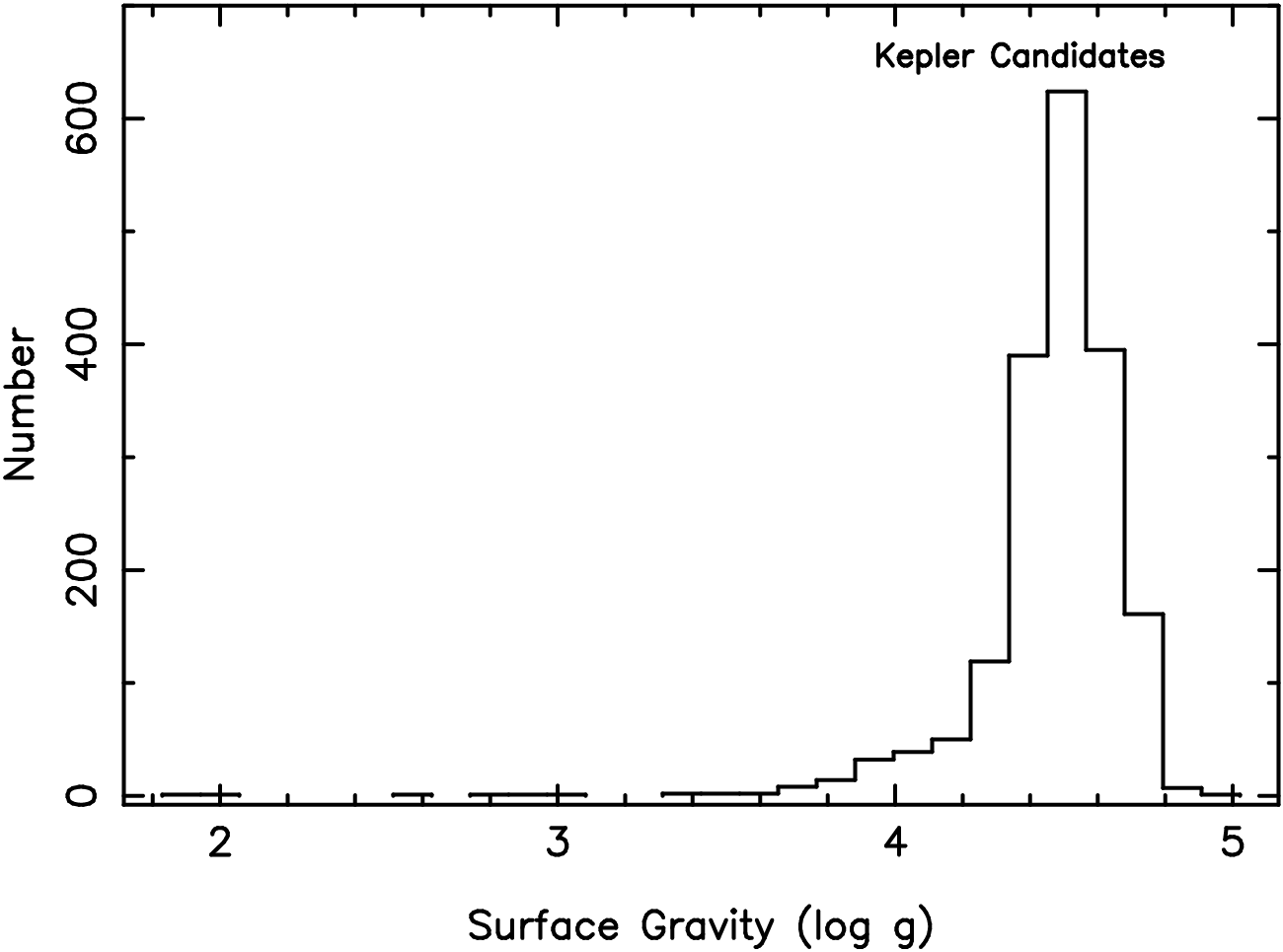} &
      \includegraphics[angle=270,width=8.2cm]{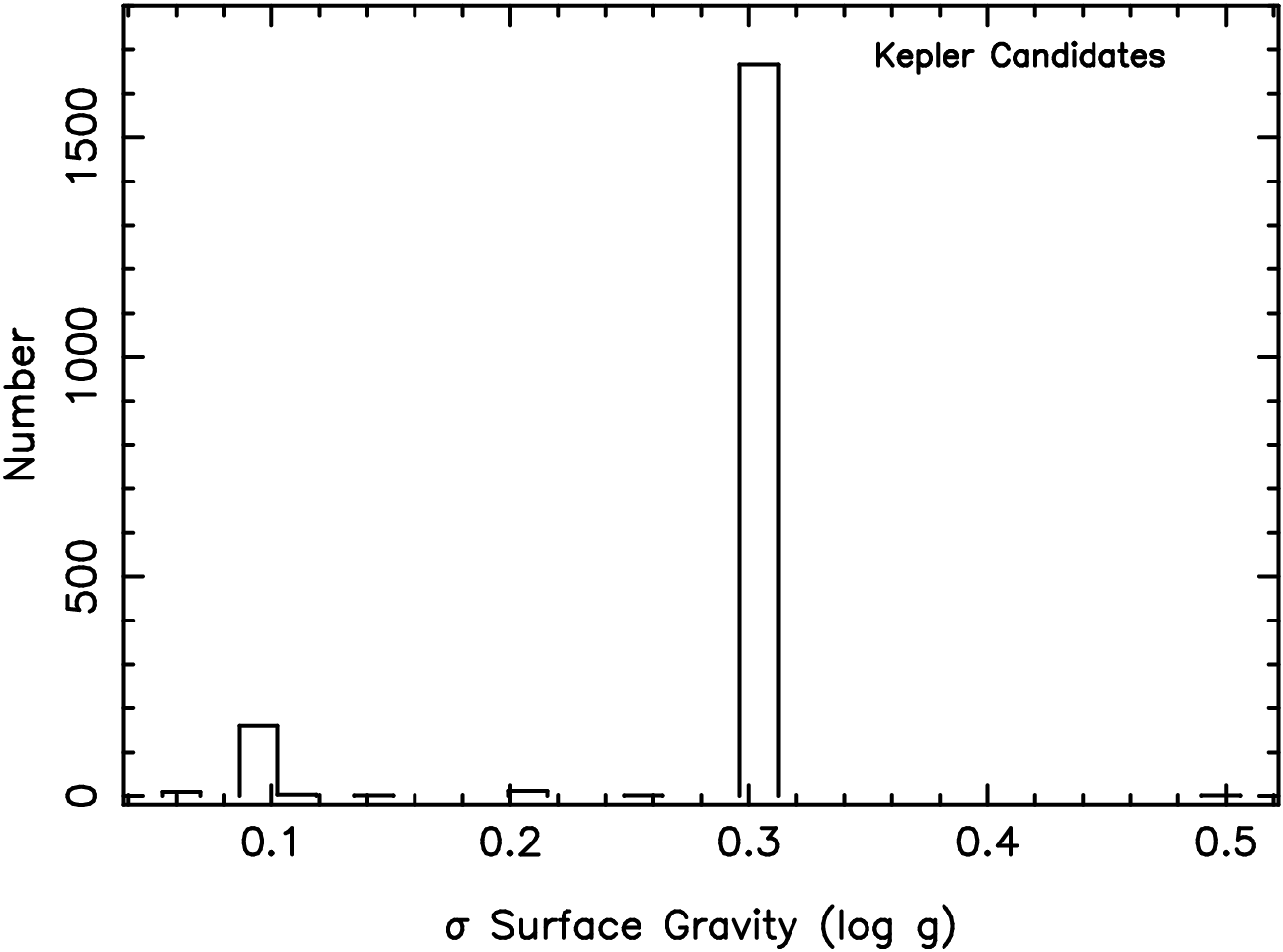}
    \end{tabular}
  \end{center}
  \caption{Histograms of the stellar parameters (left column) and
    their associated uncertainties (right column) for the Kepler
    candidates group. Included parameters are effective temperature
    (top row), stellar radius (middle row), and surface gravity
    (bottom row). Each histogram uses a total of 30 bins with data for
    1852 host stars. Note that the stellar radius histograms have a
    logarithmic scale.}
  \label{kcparams}
\end{figure*}

\begin{figure}
  \includegraphics[angle=270,width=8.2cm]{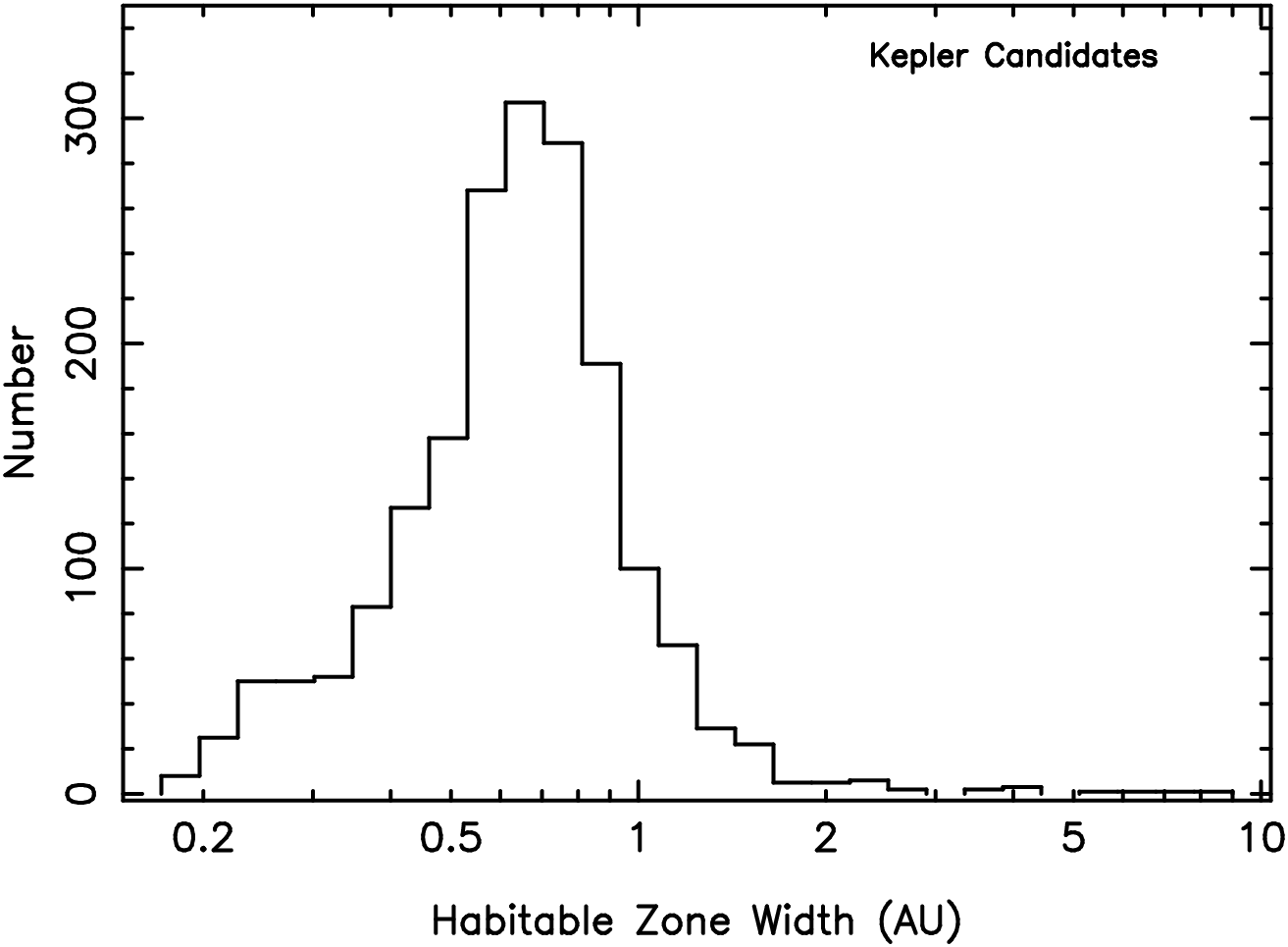} \\
  \includegraphics[angle=270,width=8.2cm]{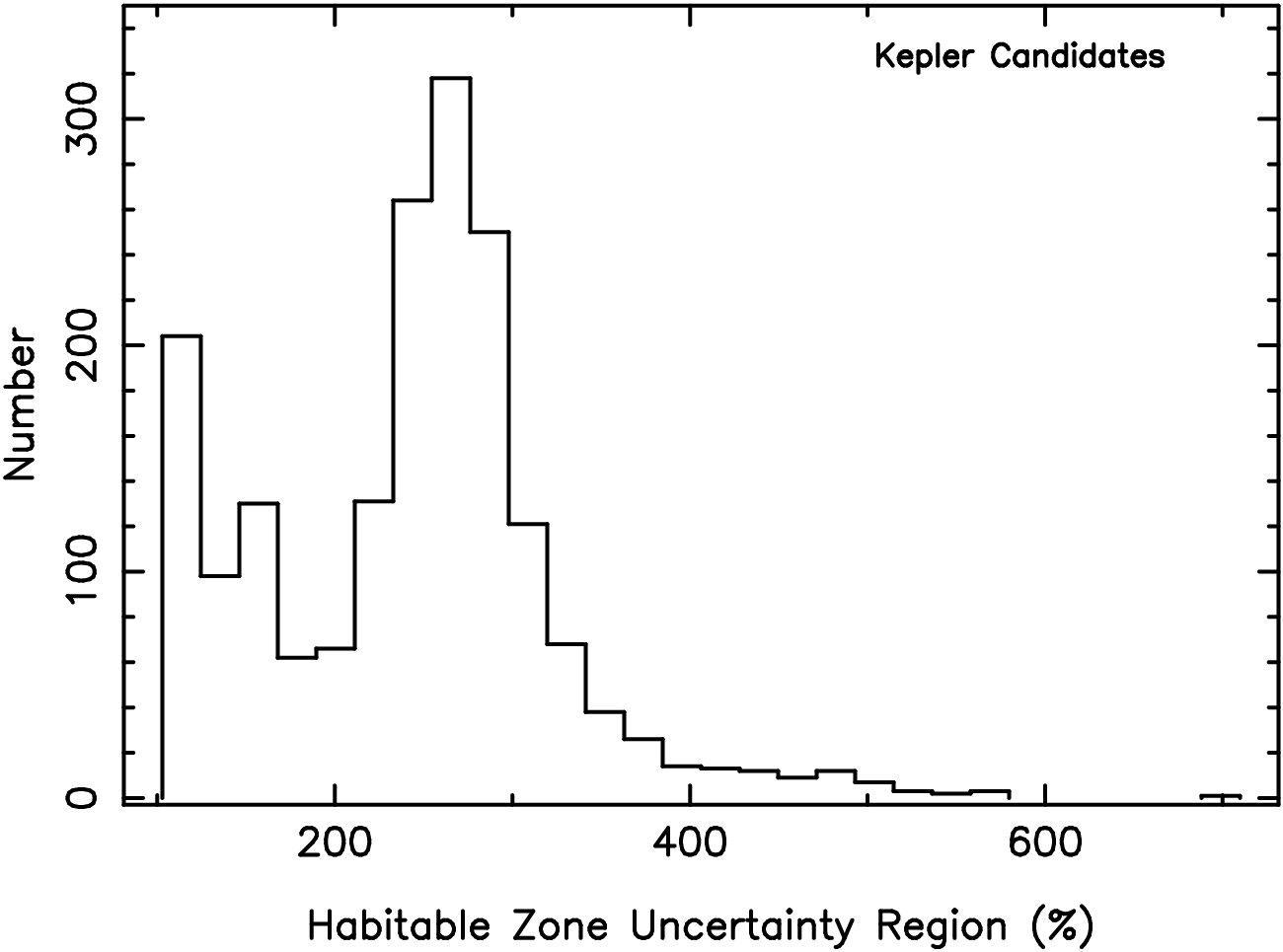}
  \caption{Histogram representations of the HZ distributions for the
    Kepler candidates group. Top: Histogram of the calculated HZ width
    (AU). Bottom: Histogram of the width of the HZ uncertainty region,
    calculated as a percentage of the HZ width shown in the top
    panel.}
  \label{kchz}
\end{figure}

The number of Kepler candidate host stars which meet our criteria of
available stellar parameters is 1852. The distributions of the
$T_\mathrm{eff}$, $R_\star$, and $\log g$ and their associated
uncertainties are shown in the histograms of Figure \ref{kcparams}.
As with the confirmed exoplanets group (see Figure \ref{cpparams})
each of the histograms are divided into 30 bins. However, the y-axes
of the surface gravities have a different scale to the other
histograms due to their substantially different distributions. A
general difference that can be seen between the confirmed exoplanet
and Kepler candidate host stars is that the uncertainty distributions
for the Kepler candidate host stars are skewed towards higher
uncertainties. This is not unexpected since the Kepler stars are
systematically fainter than those monitored by most ground-based RV
and transit programs. This does however have a significant effect on
HZ calculations as we will soon show.

The distribution of $T_\mathrm{eff}$ for the Kepler candidate host
stars is similar to that for the confirmed exoplanet host stars but
with a strong emphasis on solar (G-type) stars. The uncertainty
distribution peaks at $\sim 80$~K, almost twice that of the confirmed
exoplanet host stars.

The stellar radii distribution shown in Figure \ref{kcparams} is
likewise similar to that shown in Figure \ref{cpparams} with
relatively few giant stars in the sample. However, the uncertainty
distribution is significantly worse with a major peak in the bimodal
distribution of $\sim 0.4 R_\odot$, an order of magnitude higher than
that of the confirmed exoplanet host stars. As stated earlier, large
efforts have been made to better characterize the Kepler stars,
particularly the exoplanet candidate host stars, resulting in a minor
distribution with radius uncertainties less than $\sim 1 R_\odot$.

Determining $\log g$ values for the relatively faint Kepler stars is a
difficult task, and several efforts have been made to do so from
photometric calibrations \citep{cla11,cre13}. The histogram of $\log
g$ values shown in Figure \ref{kcparams} is consistent with most of
the Kepler candidate host stars being selected because of their dwarf
classification and thus consistent with the $T_\mathrm{eff}$ and
$R_\star$ distributions. The uncertainties for $\log g$ are
particularly unreliable with most having been assigned a default value
of 0.3~dex \citep{bro11}. This does not affect the HZ calculations
presented here since we use the $R_\star$ values and uncertainties due
to their availability.

The widths of the conservative HZ for all of the Kepler candidate host
stars are shown in the top panel of Figure \ref{kchz}. These were
calculated in the same way as for the confirmed exoplanet host
stars. An important difference between the two is that the
distribution of HZ widths for the Kepler candidate host stars peaks at
$\sim 0.7$~AU, compared with $\sim 0.9$~AU for the confirmed exoplanet
host stars. This indicates that the HZ boundaries are also closer to
the Kepler host stars than for the confirmed exoplanet host stars. The
$T_\mathrm{eff}$ distribution is similar between the two groups so it
is the slightly smaller radii distribution of the Kepler candidate
host stars that results in the overall reduction in HZ widths. The
main difference between the two groups can be seen in the bottom panel
of Figure \ref{kchz}. The slight bimodal distribution in the stellar
radii results in a similar bimodal distribution in the width of the HZ
uncertainty region. However, the majority of the Kepler stars have a
HZ uncertainty region which is between 200--400\% larger than the
width of the HZ; significantly worse than for the confirmed exoplanet
host stars. The repercussion of this is that Kepler candidate
exoplanets whose orbits are supposed to lie within the HZ of their
host star are unlikely to fall within the HZ at all. Thus statistics
of Kepler HZ candidates must be individually examined to determine if
they reasonably qualify to retain their HZ status.


\section{Specific Habitable Zone Systems}

Here we consider some specific systems to study the extent of the HZ
both with and without stellar parameter uncertainties accounted for.

The GJ~581 system has been of particular interest with regards to the
HZ boundaries as there have been claims that planets within the system
have HZ status \citep{vog10}. The host star is especially well
characterized due in no small part to the long-baseline
interferometric observations carried out by \citet{von11}. These
measurements resulted in determining the fundamental stellar
parameters of $T_\mathrm{eff} = 3498 \pm 56$~K and $R_\star = 0.299
\pm 0.010 R_\odot$.

\begin{figure}
  \includegraphics[angle=270,width=8.2cm]{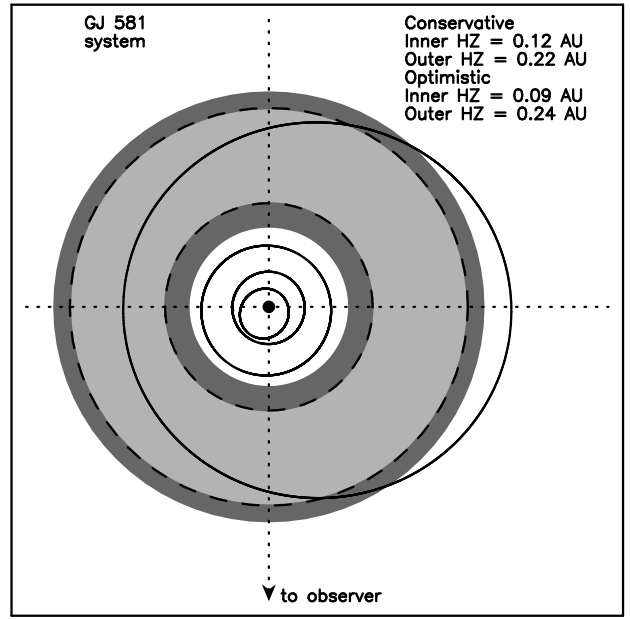}
  \caption{The calculated extent of the conservative (light-gray) and
    optimistic (dark-gray) HZ for the GJ~581 system. The Keplerian
    orbits of the planets are shown as solid lines. The 1$\sigma$
    uncertainty boundaries for the conservative HZ are indicated by
    dashed lines.}
  \label{gj581}
\end{figure}

Figure \ref{gj581} shows a top-down view of the GJ~581 system with the
calculated HZ regions. The light gray represents the conservative HZ
model and the dark gray regions represent the extensions to the HZ
from the optimistic model. The orbits of the GJ~581 planets are
overlaid on the plot (solid lines) where we have adopted the
four-planet model described by \citet{for11}. The dashed lines
indicate the 1$\sigma$ extensions of the conservative model boundaries
due to the stellar parameter uncertainties. In this case, these
1$\sigma$ uncertainties are almost negligible in size and the HZ
uncertainty region (see Figures \ref{cphz} and \ref{kchz}) is 102\% of
the HZ width of 0.104~AU. Thus, the location of the HZ is well-defined
for this system.

\begin{figure}
  \includegraphics[angle=270,width=8.2cm]{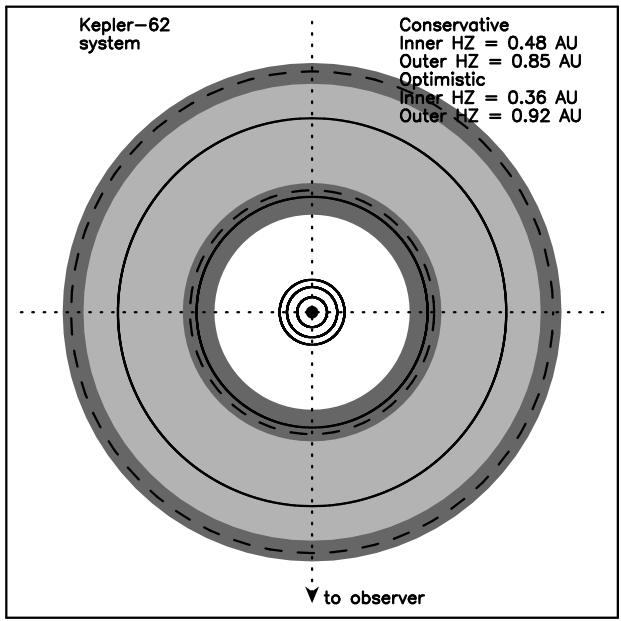}
  \caption{The calculated extent of the conservative (light-gray) and
    optimistic (dark-gray) HZ for the Kepler-62 system. The Keplerian
    orbits of the planets are shown as solid lines. The 1$\sigma$
    uncertainty boundaries for the conservative HZ are indicated by
    dashed lines.}
  \label{kepler62}
\end{figure}

\begin{figure}
  \includegraphics[angle=270,width=8.2cm]{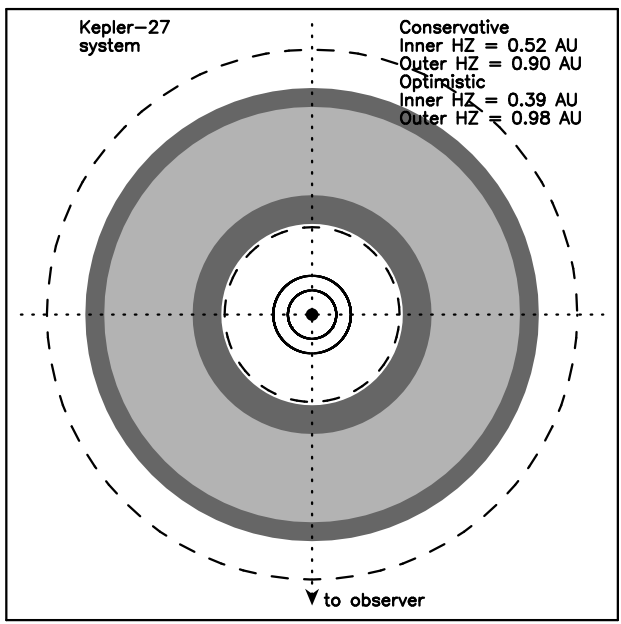}
  \caption{The calculated extent of the conservative (light-gray) and
    optimistic (dark-gray) HZ for the Kepler-27 system. The Keplerian
    orbits of the planets are shown as solid lines. The 1$\sigma$
    uncertainty boundaries for the conservative HZ are indicated by
    dashed lines.}
  \label{kepler27}
\end{figure}

A prominent confirmed multi-planet system detected by the Kepler
mission is that of Kepler-62 \citet{bor13}. This five-planet system
includes two which early estimates showed are in the HZ of the host
star. The published stellar parameters are $T_\mathrm{eff} = 4925 \pm
70$~K and $R_\star = 0.64 \pm 0.02 R_\odot$. The HZ conservative and
optimistic regions are shown in Figure \ref{kepler62}. As noted by
\citet{bor13}, the Kepler-62 system is unstable with the Keplerian
orbital parameters presented in their paper due to overlapping
orbits. We therefore adopt circular orbits for the planets which are
overlaid on the plot. The width of the conservative HZ region is
0.37~AU. The dashed lines for the 1$\sigma$ uncertainty boundaries
show that the HZ location is less well-known in this case than it is
for GJ~581. The HZ uncertainty region for Kepler-62 is 120\% of the HZ
width. Even so, the uncertainties are small enough such that the e and
f planets are likely to be in the optimistic and conservative HZ
regions as shown. To quantify this, we assume a normal distribution of
the HZ boundary uncertainties. The e planet is 1.92$\sigma$ away from
the conservative HZ boundary and thus 94.5\% likely to be outside of
this region. The f planet is 2.77$\sigma$ away from the conservative
HZ boundary and thus 99.4\% likely to be inside of the conservative HZ
region.

The two planets of the Kepler-27 system were confirmed using the
Transit Timing Variation (TTV) technique by \citet{ste12}. Although
neither of these planets are purported to lie within the HZ, this is
an interesting system to study as an example of one with relatively
large stellar parameter uncertainties. For Kepler-27, these are
$T_\mathrm{eff} = 5400 \pm 60$~K and $R_\star = 0.59 \pm 0.15
R_\odot$. The HZ and planets for this system are shown in Figure
\ref{kepler27}. The width of the conservative HZ is 0.38~AU and the HZ
uncertainty region is 201\% of this width. As described in Section
\ref{application}, an uncertainty region of this size results in
considerable doubt as to the HZ status of any planet described to be
in HZ in such cases. We have shown that the vast majority of the
Kepler exoplanet candidate fall into this category with the currently
known stellar parameters.


\section{Conclusions}

The HZ is an increasingly important property of exoplanet host stars
with the ever-increasing sensitivity to exoplanets of smaller
size/mass and at longer orbital periods. The greatest hindrance to
understanding the properties of exoplanets is the difficulty in fully
characterizing the host star properties. This also results in
limitations in defining the extent of the HZ in exoplanetary systems.

Here we have shown these limitations as imposed by the specific
stellar properties of $T_\mathrm{eff}$, $R_\star$, and $\log g$. The
importance of $R_\star$ is in deriving the stellar luminosity and the
value of $\log g$ is utilized only when the radius is not determined
through other means. The stellar parameters for the confirmed
exoplanet host stars are sufficiently well determined such that the
size of the HZ uncertainty region lies below 150\% for most of the
stars. However, the HZ uncertainty region distribution for the Kepler
candidate host stars is dominated by those in the 200-400\% range
where the HZ status of exoplanets is highly dubious. Analysis of the
stellar properties between the two groups shows that the uncertainty
in stellar radius is the primary cause of this HZ uncertainty
difference.

With the continuous rate of new discoveries from both ground and
space-based surveys, the search for terrestrial-size planets in the HZ
of their host stars is a difficult but achievable goal. When such
discoveries are made, it will always be critical to quantify the
extent to which we can correctly classify these discoveries as HZ
planets. Without such analysis, the understanding of the frequency of
Earth-size planets in the HZ (sometimes referred to as $\eta_\oplus$)
will be less secure than we suppose.


\section*{Acknowledgements}

The author would like to thank Daniel Huber for suggesting this
investigation and providing feedback on the manuscript. The author
would also like to thank Ravi Kopparapu and the anonymous referee for
productive feedback. This research has made use of the following
archives: the Exoplanet Orbit Database and the Exoplanet Data Explorer
at exoplanets.org, the Habitable Zone Gallery at hzgallery.org, and
the NASA Exoplanet Archive, which is operated by the California
Institute of Technology, under contract with the National Aeronautics
and Space Administration under the Exoplanet Exploration Program.


\end{document}